\documentclass{article}

\usepackage{subfig}
\usepackage{graphicx}
\usepackage{color}
\usepackage{soul}
\usepackage{bm}
\usepackage{setspace}
\usepackage[right]{lineno}

\title{Limitations of Time Resources in Human Relationships Determine Social Structures}

\author{Masanori Takano$^1$ \& Ichiro Fukuda$^1$\\
$^1$ CyberAgent, Inc., Chiyoda-ku, Tokyo, Japan}

\begin{document}
\maketitle

\begin{abstract}
{\bf The number of social relationships that a single human being can possibly be involved in is limited because individuals face time constraints (that is, time costs) in constructing and maintaining social relationships; 
further, the distribution of the strength of such relationships (as measured by frequency of social interaction) looks significantly skewed (a power law distribution), that is, a few strong relationships and many weak relationships.
This skewedness suggests that the costs and benefits of bonding with others depend on the strength of the social relationships: if it involves uniform costs and benefits, the distribution would not be skewed. 
The bonding is known as social grooming.
That is, humans strategically construct their social relationships, and thus, complex human societies should also be strategically constructed. 
Therefore, it is important to know their strategies for understanding human societies. 
Previous studies provide evidence of social grooming strategies by examining the evolution of social grooming methods and the difference between the various social grooming methods. 
However, quantitative laws that are important for theoretically understanding human societies are still open to investigation.
Social big data is a particularly powerful tool for finding such laws.
Therefore, we analyse data from six communication systems (Twitter, a social networking site providing two types of interactions, an avatar chat, a mobile phone and a short message service). 
We find a simple quantitative law by which social relationships are constrained $Nm^a$ ($a > 1$); 
here, $N$ is the number of social relationships and $m$ is a mean of the strength of those relationships. 
The fact that deep social relationships require higher costs per relationship than shallow relationships is suggested by $a > 1$ (if the both is equal then $a$ will be $1$), because the effect of $m$ on the constraint increases with $m$. 
For exploring why $a$ is greater than $1$, we conduct an individual-based simulation where social grooming costs are assumed to increase linearly with the strength of social relationships. 
Our results indicate that this model fits all data sets, that is, it displays an explanation capacity for the phenomenon. 
In addition, an analysis of this simulation proves our assumption about the social grooming cost increasing with the strength of social relationships as being true. 
Moreover, it suggests that its gradient increases the width and shallowness of these relationships. 
The law and its causes suggest that mankind's evolution of social grooming has enabled changing social structures, and the phenomenon is due to the constraints of the social network generation.
These findings will contribute towards an explanation of the evolution of the various social grooming methods of humans and their significantly large social group. }
\end{abstract}

Based on the social brain hypothesis~{\bf(Byrne R and Whiten A, 1989; Whiten A and Byrne R, 1997; Dunber RIM, 2000; Dunbar RIM, 2003)},
an explanation of social structures, which are typical for humankind, is expected to offer knowledge about human origins, such as the limitation on the number of social relationships~{\bf(Dunber RIM, 2000; Dunbar RIM, 2003; Hill RA and Dunbar RIM, 2003; Gon{\c{c}}alves B et al., 2011; Arnaboldi V et al., 2013a)}
and the skewness of social relationships~{\bf(Zhou WX et al., 2005; Hossmann T et al., 2011; Song C et al., 2012; Arnaboldi V et al., 2012; Hu T et al., 2013; Arnaboldi V et al., 2013b; Fujihara A et al, 2014)},
that is, its distributions following a power law~{\bf(Song C et al., 2012; Pachur T et al., 2012; Arnaboldi V et al., 2012; Fujihara A et al, 2014; Hossmann T et al., 2011; Hu T et al., 2013)}.
These social structures may be caused by human beings' social grooming strategies~{\bf(Dunbar RIM, 2004)}.

Social grooming is used to construct and maintain social relationships. 
This behaviour is important in complex human societies~{\bf(Dunber RIM, 2000)}. 
Close social relationships lead to mutual cooperation~{\bf(Uchino BN et al., 1996; Silk JB et al., 2003; Hill RA et al., 2008; Russell YI and Phelps S, 2013; Takano M et al., 2016a; Takano M et al., 2016b; Dunbar RIM, 2016)}. 
A high sociability among baboon mothers, for example, increases the probability of their children's survival, thanks to increased cooperation from others~{\bf(Silk JB et al., 2003)}; 
similarly, humans tend to co-operate more with close friends than with others~{\bf(Takano M et al., 2016a; Dunbar RIM, 2016)}. 
On the other hand, having many weak social relationships helps in obtaining a variety of information which is advantageous in a complex society~{\bf(Granovetter M, 1973; Dunbar RIM, 2004; Eagle N et al., 2010; Arnaboldi V et al., 2013c)}. 

Social relationships provide humans with various advantages. 
However, they face cognitive constraints~{\bf(Dunbar RIM, 2012)} 
(for example, memory and processing capacity) and time constraints (that is, time costs) in constructing and maintaining social relationships. 
These time costs are not negligible, as humans spend a fifth of their day in social grooming~{\bf(Dunbar RIM, 1998)} 
and maintaining social relationships~{\bf(Hill RA and Dunbar RIM, 2003; Roberts SGB and Dunbar RIM, 2011)}. 
Therefore, the mean strength of existing social relationships has a negative correlation with the number of social relationships~{\bf(Roberts SGB et al., 2009; Miritello G et al., 2013a)}. 

Humans must construct and maintain social relationships within the constraints of this trade-off. 
Thus, we expect that they employ strategies to distribute the limited time resources to maximize benefits from their social relationships~{\bf(Brown SL and Brown RM, 2006; Miritello G et al., 2013b; Saramaki J et al., 2014)}. 
As a result of such strategies, social relationship strengths (as measured by frequency of social grooming~{\bf(Roberts SGB and Dunbar RIM, 2011; Arnaboldi V et al., 2012; Song C et al., 2012; Arnaboldi V et al., 2013b; Fujihara A et al, 2014; Saramaki J et al., 2014)} 
may often show a much skewed distribution~{\bf(Zhou WX et al., 2005; Arnaboldi V et al., 2013b)} 
(distributions following a power law~{\bf(Song C et al., 2012; Arnaboldi V et al., 2012; Fujihara A et al, 2014; Hossmann T et al., 2011; Hu T et al., 2013)}). 

In other words, human beings' time cost distribution strategies should affect their structures of social relationships. 
Similarly, the advantage of strategies may depend on social grooming methods because recipients' satisfaction levels depend on social grooming methods, that is, face to face and video call communications show higher satisfaction levels than phone and text communications~{\bf(Vlahovic T et al., 2013)}, 
and their time and efforts involved are also different. 
Humans have various social grooming methods, such as physical contacts~{\bf(Nelson H, 2007)},
gaze grooming~{\bf(Kobayashi H and Kohshima S, 1997; Kobayashi H and Hashiya K, 2011)}, 
gossip~{\bf(Dunbar RIM, 2004)}, 
greetings, phone calls~{\bf(Vlahovic T et al., 2013)}, 
e-mails~{\bf(Vlahovic T et al., 2013)} 
and social networking sites (SNS)~{\bf(Burke M and Kraut RE, 2014; Scissors L et al, 2016)}. 
These methods will have differing effects in the formation of social relationships and on the time costs.

Human beings maintain large complex social groups by using various and effective social grooming methods efficiently~{\bf(Dunbar RIM, 2004)}. 
Humans have acquired these methods during the evolutionary process. 
Apes, which are closely related to humans, clean each other's fur as social grooming~{\bf(Nakamura M, 2003)}. 
In contrast, humans almost do not do grooming~{\bf(Nelson H, 2007)}. 
Alternatively, human beings do lightweight social grooming which has evolved to adapt large groups, for example, gaze grooming~{\bf(Kobayashi H and Kohshima S, 1997; Kobayashi H and Hashiya K, 2011)} 
and gossip~{\bf(Dunbar RIM, 2004)} 
that enable humans to have several social relationships and less time and efforts are required for these methods.
These previous studies provide evidence of social grooming strategies by examining the evolution of social grooming methods. 
However, quantitative laws that are important for theoretically understanding human societies are still open to investigation.
Additionally, humans often invest in novel social grooming methods on the Internet, such as social networking sites, online chats and video calls. 
We will need to know the effects of social grooming methods on social relationships for a safe and comfortable relationship. 
However, little is known about its effects and mechanisms~{\bf(Arnaboldi V et al., 2013c)}. 

This study aims to discover how social grooming methods influence time cost distribution strategies of the methods and social structures depending on those strategies. 
We analyse the following six data sets of communication systems: 
Twitter data (used as a test set in~{\bf(Cheng Z et al., 2010)}, 
755 group chats (see the supplementary information (SI) Fig. 1 for specifications), 755 wall communications (see S1 Fig. for specifications), Japanese avatar chat Ameba Pigg (see SI Fig. 2 for specifications), mobile phone calls~{\bf(Madan A et al, 2012)}, 
and short message service (SMS)~{\bf(Madan A et al, 2012)}. 
These human behaviour data sets enable us to analyse social interactions among several users quantitatively. They are useful for constructing theoretical models of social phenomena. 
Studies~{\bf(Arnaboldi V et al., 2012; Song C et al., 2012; Fujihara A et al, 2014; Hossmann T et al., 2011; Arnaboldi V et al., 2013b; Hu T et al., 2013; Takano M et al., 2016a; Saramaki J et al., 2014; Dunbar RIM, 2016; Takano M et al., 2016b)} 
previously cited in this paper have also analysed data from SNSs, mobile phones, social network games, and SMSs.

\section*{Data Analysis}

\subsection*{Data Sets}
We used six data sets (see table~\ref{tbl_dset} for details): 
1) Twitter data (used as test set in~{\bf(Cheng Z et al., 2010)} 
recording interactions among $2,585$ people with $278,475$ relationships, from 23/6/2007 to 17/3/2010, where an act of social grooming was defined as using the ``mention'' or ``reply'' functions to communicate with others; 
2) and 3) Data from the Japanese SNS 755, published by 7gogo, Inc. (http://7gogo.jp/), which provides two types of communication systems data, dating from 1/1/2015 to 31/3/2015, which we treated as two different sets (see SI Fig. 1 for specifications), namely data from group chats and that from wall communications. The former data records interactions among $17,796$ users with $238,611$ relationships, where we defined an act of social grooming as communicating in a chat limited to two members. 
The latter data records interactions among $20,000$ users with $534,475$ relationships, where we defined an act of social grooming as posting a comment on another's wall. 
We removed data relevant to official users from both data sets; 
4) Data from Japanese avatar chat Ameba Pigg, published by CyberAgent, Inc. (https://pigg.ameba.jp/), which records interactions among $76,379$ users with $1,610,710$ relationships, from 1/10/2014 to 31/12/2015 (see SI Fig. 2 for specifications), where we defined an act of social grooming as communicating in a chat limited to two members; 
5) Data from mobile phone calls~{\bf(Madan A et al, 2012)}, 
recording mobile phone calls among $73$ people with $7,805$ relationships from 5/9/2008 to 29/6/2009, where we defined an act of social grooming as a call to another; 
6) Data from SMS~{\bf(Madan A et al, 2012)}, 
which records SMSs among $61$ people with $2,266$ relationships from 1/1/2008 to 27/6/2009, where we defined an act of social grooming as sending a message to another.

In the data sets from Twitter, 755 (group chat and wall communication) and Ameba Pigg, we limited the targets of analysis to active users who had more number of social grooming days than the $50$th percentile among Twitter users and the $75$th percentile among 755 and Ameba Pigg users because these internet service data sets included many inactive users.
In this paper, we defined the strength of social relationships $d_{ij}$ as the days on which individual $i$ does social grooming to individual $j$.

\subsection*{Analysis}

\begin{figure}[ht]
  \begin{center}
\includegraphics[width=0.8\columnwidth]{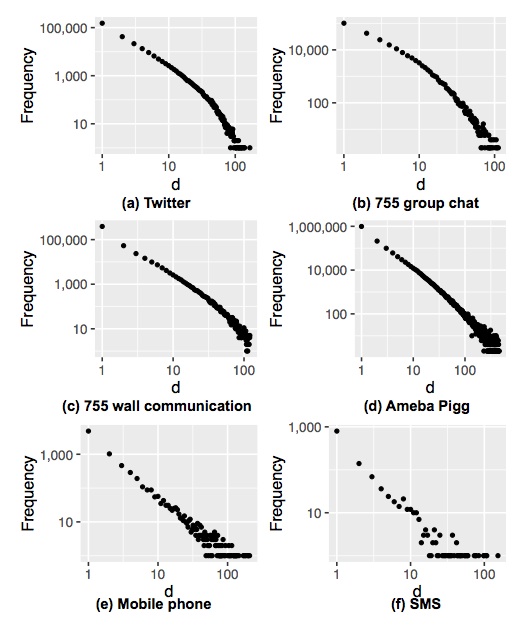}
  \caption{
  Power law distributions of number of days $d_{ij}$, where individual $i$ interacted with individual $j$.
  These results were similar to those of previous studies~{\bf(Hossmann T et al., 2011; Song C et al., 2012; Arnaboldi V et al., 2012; Hu T et al., 2013; Fujihara A et al, 2014)}. 
  Power law coefficients were $1.92$ (A: Twitter), $3.71$ (B: 755 group chat), $2.29$ (C: 755 wall communication), $1.97$ (D: Ameba Pigg), $1.98$ (E: mobile phone) and $2.00$ (F: SMS).}
  \label{fig_powerlaw}
\end{center}
\end{figure}

\begin{figure}[ht]
  \begin{center}
\includegraphics[width=0.6\columnwidth]{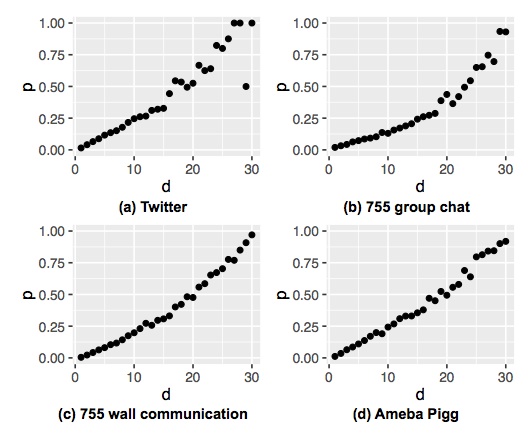}
  \caption{
  The Yule--Simon process on social grooming strategies.
  These figures show probability $p$ of social grooming on days after each strength of social relationship $d$.
  This indicates that the power law distributions were generated by the Yule–-Simon process because the $p$ was proportional to $d$, and these strategies subsequently generated the power law distributions.
  The data periods were from the first thirty days. 
  We did not observe the similar trends in the mobile phone and SMS data sets due to insufficient data.
  These results were similar in a previous study~{\bf(Pachur T et al., 2012)}. 
  }
  \label{fig_yule}
\end{center}
\end{figure}

In the six communication systems examined in this study, the distributions of the strengths of social relationships $d$ showed power law distributions (Fig.~\ref{fig_powerlaw}), similar to those observed in previous studies~{\bf(Arnaboldi V et al., 2012; Song C et al., 2012; Fujihara A et al, 2014; Hossmann T et al., 2011; Hu T et al., 2013)}. 
Individuals selected social grooming partners in proportion to the strength of their social relationships, that is, the individuals tended to reinforce their strong social relationships; these power law distributions were generated by the Yule–-Simon process~{\bf(Yule GU, 1925; Simon HA, 1955; Newman MEJ, 2005)} 
(Fig.~\ref{fig_yule}), similar to a previous study~{\bf(Pachur T et al., 2012)}. 
Thus, individuals distributed limited time resources in proportion to the strength of their social relationships and this led to the further strengthening of these relationships. 
It shows that individuals selected social grooming partners depending on the strength of their social relationships.

As above, these skew distributions seem to be caused by individuals' time cost distribution strategies. 
Here, we considered a simple model as a null model to analysed how individuals construct social relationships depending on their strength under time cost constraints.
If the daily social grooming cost is independent of the strength of the relationship then an individual's total social grooming cost $C$ is $\sum_{i=1}^N d_i$, where $d_i$ is the strength of social relationships from the individual to individual $i$, and $N$ is the total number of social grooming partners. 
If $m$ is the mean of strengths of social relationships (that is, $\sum_{i=1}^N d_i/N$), then $C = \sum_{i=1}^N d_i = Nm$. 
Therefore, under this assumption, $N$ should be inversely proportional to $m$. 
We conducted the following regression analysis to confirm this hypothesis in the data sets:
\begin{eqnarray}
\log N &\sim& Normal(\mu, \sigma) \label{eq_cmn_reg}, \\
\mu &=& -a \log m + b \log u, \nonumber
\end{eqnarray}
where $u$ is the number of days of participation for each user, that is, we assumed that a user's total social grooming costs were equal to the number of days for which he/she had participated in the activity ($C = u^b$).
If this hypothesis is correct, then $a$ (the coefficient of $\log m$) should be $1$.

\begin{figure}[ht]
\begin{center}
\includegraphics[width=0.95\columnwidth]{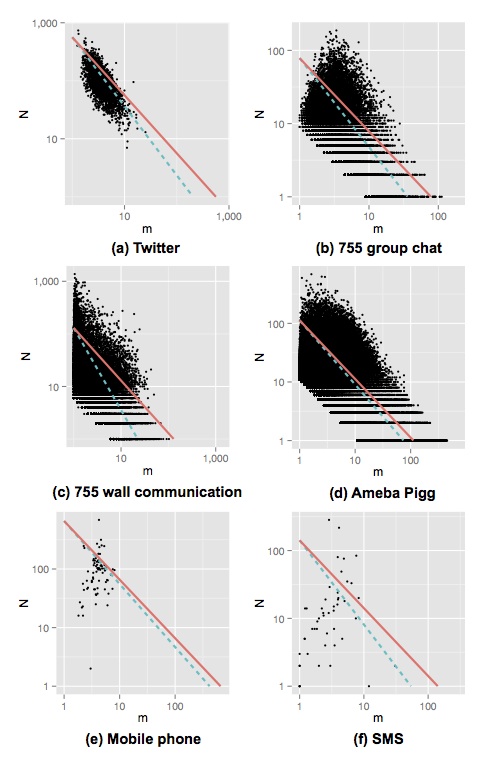}
\caption{
A simple law between $N$ (total number of social grooming partners) and $m$ (mean number of social grooming days).
These user behaviour data (black points) did not obey the null model ($C = Nm$; orange lines), where the assumed social grooming cost was independent of the strength of the relationship, but obeyed $C = Nm^a$ (green and dash lines), where $a =1.19$ (A: Twitter), $a = 1.21$ (B: 755 group chat), $a =1.56$ (C: 755 wall communication), $a =1.10$ (D: Ameba Pigg), $a =1.07$ (E: mobile phone) and $a =1.24$ (F: SMS), where user use-days $u$ were $75$th percentile, estimated by the regression models (Eq. \ref{eq_cmn_reg}), where adjusted R-squares were $0.990$ (A: Twitter), $0.974$ (B: 755 group chat), $0.959$ (C: 755 wall communication), $0.997$ (D: Ameba Pigg), $0.994$ (E: mobile phone), $0.990$ (F: SMS) (see table \ref{tbl_pt_nb} for details), and $a>1$ were significant excluding the mobile phone data set.
}
\label{fig_cmn}
\end{center}
\end{figure}

Figure \ref{fig_cmn} shows that all data sets obey $C = Nm^a$ ($a > 1$) (see table~\ref{tbl_pt_nb} for the regression results).
That is, the null model did not fit the data sets, that is, social grooming cost is not independent of the strength of the relationship.
In other words, individuals who had a few strong relationships (that is, large $m$) invested more in relationships than individuals who had many weak relationships (that is, small $m$).
$a > 1$ shows that for $C = Nm^a$, the stronger the social relationship $d$, the more social grooming costs increased per day because the effect of $m$ on $C$ increases with $m$.

\begin{figure}[ht]
\begin{center}
\includegraphics[width=0.95\columnwidth]{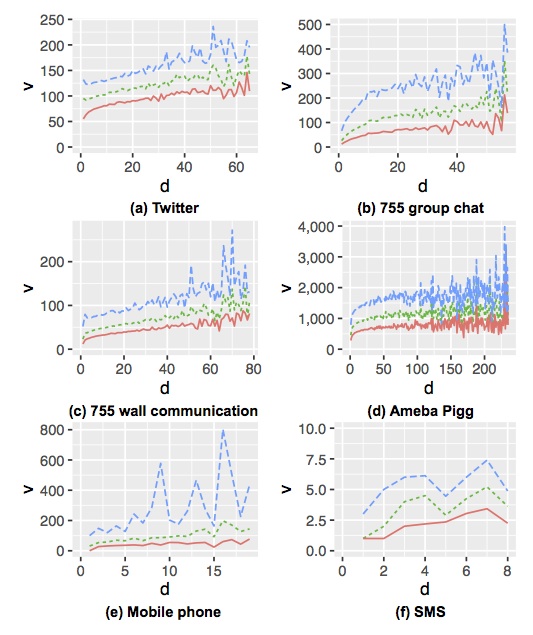}
  \caption{
  Increasing communication volumes per day $v$ by strengths of social relationships $d$,
  where the volumes were number of characters per day in Twitter, 755 and Ameba Pigg, duration per day in mobile phones and frequency of messaging per day in SMS because we did not consider any information regarding number of characters in the SMS data set.
  The orange lines are the $25$th percentile, the green and dotted lines are the $50$th percentile and the blue and dashed lines are the $75$th percentile. 
  These are shown for cases where the number of samples was more than $20$ (the ranges of the $d$ of mobile phone and SMS are short because these were smaller data sets). 
  }
\label{fig_strlen}
\end{center}
\end{figure}

\begin{figure}[ht]
  \begin{center}
\includegraphics[width=0.7\columnwidth]{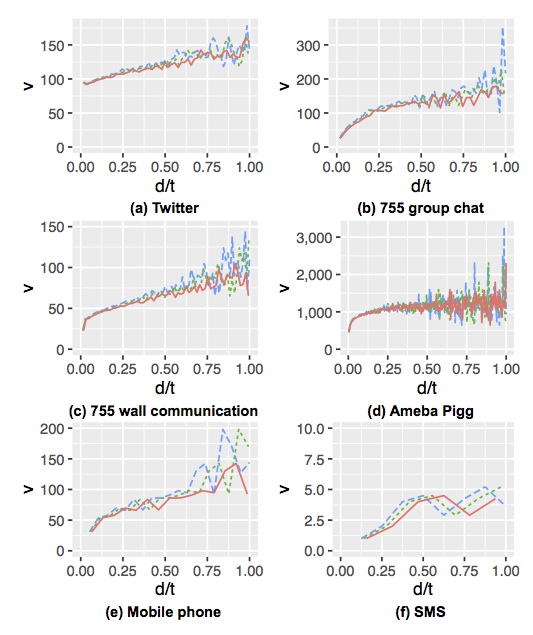}
\caption{
  The gradients of communication volumes depending on social grooming density as distinct from those depending on social grooming frequency.
  These figures show compaction of the medians of $v$ for each social grooming density ($d/t$) for different periods ($t$ is number of days of the data periods).
  Each line represents entire periods (orange lines), nine-tenths of the periods (green and dotted lines), and eight-tenths of the periods (blue and dashed lines).
  These are shown when the number of samples is more than $20$ (the ranges of $d$ of mobile phone and SMS are short because the sizes of these data sets were small). 
 }
  \label{fig_com_vol_comp}
\end{center}
\end{figure}

To determine how the strength of relationship $d$ affects social grooming costs, we analysed the relation between communication volumes $v$ and the strength of social relationships $d$. 
Figure \ref{fig_strlen} shows that the strengths increased the volumes of communication per day, that is, social grooming costs also probably increased with increase in strengths.
Figure \ref{fig_com_vol_comp} shows that the increase in gradients was analogized in each data set, that is, this shows gradients depending on social grooming density as distinct from those depending on social grooming frequency.
That is, the results of the analysis show that communication volumes $v$ increased along with the relationship strength $d$ and the gradients were independent of the number of days of the data periods $t$. 
Consequently, social grooming cost should increase with an increase in social grooming density ($d/t$), under the assumption that social grooming costs are proportional to communication volume $v$.

\section*{Individual-based Simulation}

\subsection*{Model}
To explain $a > 1$ for $C = Nm^a$, we constructed a simulation model of social grooming cost distribution strategies based on the assumption that social grooming costs increase with an increase in social grooming density $d/t$.
We used a linear social grooming cost function $c(d) = \alpha d/t + \beta$ as the simplest assumption (if $\alpha = 0$ then it is the null model). 
In the model, we considered two type individuals which were groomers and groomees, and groomers construct social relationships using their limited resources (that is, time), based on this assumption and the Yule--Simon process. 

We conducted the following simulation for $T$ step to construct social relationships $d_{ij}$. 
At each step $t$, groomer $i$ repeats the following for $R > 0$. 
$R$ is reset to an initial value $R_0$ before each step $t$.
Each $i$ has a resource $R$ that is spent when $i$ performs social grooming with others. 
Each $i$ creates a social relationship with a stranger, groomee $j$, depending on probability $q_i$, where the strength of social relationship $d_{ij}$ is 1 and $i$ pays the cost $\beta$ from its resource $R$ (if $R < \beta$, then $d_{ij}$ is $R/\beta$ and $R$ becomes $0$). 
In contrast, $i$ reinforces its social relationships depending on probability $1-q_i$. 
Each $i$ selects a social grooming partner $j$ depending on a probability proportional to the strength of the social relationships between $i$ and $j$ ($d_{ij}$), then $i$ adds $1$ to the strength of its social relationship (that is, the Yule–-Simon process) and pays the cost $c(d_{ij}) = \alpha d_{ij}/t + \beta$ from $R$ (if $R < c(d)$, then $i$ adds $R/c(d)$ to its strength of social relationship $d_{ij}$ and $R$ becomes $0$).
Each $i$ does not perform the act of social grooming twice with the same groomees in each step $t$. 
Therefore, selected groomees are excluded from the selection process of a social grooming partner $j$ in each step $t$.

\subsection*{Experiments}

\begin{figure}
\begin{center}
\includegraphics[width=0.95\columnwidth]{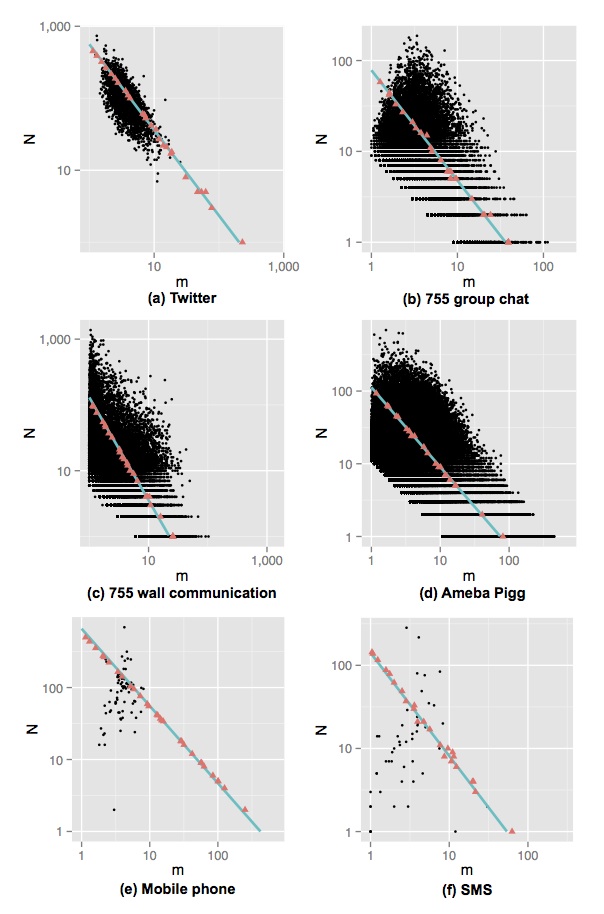}
\caption{
  Explanation capacity of the simulation model.
  The results of fits by the simulation model to the regression lines of all data sets (that is, green and dashed lines in Fig. \ref{fig_cmn}).
  Very good fits were observed between the simulation results (orange triangles) and the regression lines (green lines); 
  that is, the simulation model displayed explanation capacity for the phenomenon $a > 1$ for $C = Nm^a$, with a monotonically increasing cost function ($\alpha>0$). 
  The parameters of the cost functions were $\alpha = 1.34$ and $\beta = 0.24$ (A: Twitter), $\alpha = 1.27$ and $\beta = 0.39$ (B: 755 group chat), $\alpha = 3.89$ and $\beta = 0.23$ (C: 755 wall communication), $\alpha = 1.62$ and $\beta = 0.64$ (D: Ameba Pigg), $\alpha = 0.05$ and $\beta = 0.31$ (E: mobile phone) and $\alpha = 5.05$ and $\beta = 0.34$ (F: SMS).
}
\label{fig_cmn_sim_each}
\end{center}
\end{figure}

First, we tested the fit of this model.
That is, we optimized parameters $\alpha$ and $\beta$ to fit the regression lines of Figure \ref{fig_cmn}, where the evaluation function was the mean square error between simulation results and regression lines, $T$ was the period for each data set, $R_0$ was the $75$th percentile of each user's use-days $u$ divided by $T$ ($R_0=0.126$ and $T = 998$ (Twitter), $R_0=0.258$ and $T = 120$ (755 group chats), $R_0= 0.225$ and $T = 120$ (755 wall communications), $R_0= 0.164$ and $T = 456$ (Ameba Pigg), $R_0 = 0.589$ and $T = 297$ (mobile phone) and $R_0 = 0.107$ and $T = 543$ (SMS)). 
The number of groomees was set large enough.
If our hypothesis is correct then $\alpha > 0$.
We found that this model fitted all data sets (Fig. \ref{fig_cmn_sim_each}), that is, it had an explanation capacity for the phenomenon $a > 1$ for $C = Nm^a$, with a monotonically increasing cost function ($\alpha>0$).
It demonstrates that the phenomenon $a > 1$ can be attributed to a groomers' tendency to invest higher costs in strong rather than weak social relationships.

\begin{figure}
\begin{center}
\includegraphics[width=0.95\columnwidth]{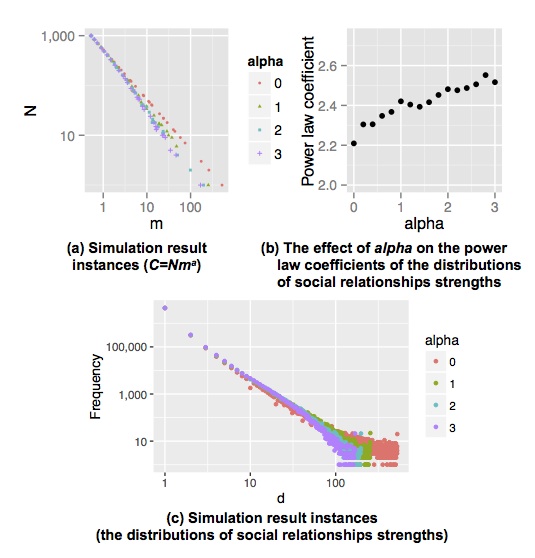}
  \caption{
  Effect of the gradient of grooming cost function $\alpha$ on $C=Nm^a$ and the distribution of the strengths of social relationships in the simulation.
  Figure A shows that gradient $\alpha$ increased $a$. 
  Figures B and C show that the gradient $\alpha$ increased the power law coefficients of the strengths of social relationships, that is, the gradient increased width and shallowness of social relationships.
  The parameters $T$, $R$ and $\beta$ were based on Twitter (see Fig. \ref{fig_cmn_sim_each}A) because $\beta$ has no effect on $a$. 
  In figures A and B, the number of individuals was $200$ and $q_i$ included the $200$ values used in $[0, 1]$. 
  In figure C, the number of individuals was $10,000$ and $q_i$ included the $10,000$ values used in $[0, 1]$.
}
\label{fig_cmn_sim}
\end{center}
\end{figure}

Next, we analysed the effect of parameter $\alpha$ on the structure of social relationships by using the model; we used $T$, $R$ and $\beta$ of the Twitter data set, as in the former experiment.
As the result demonstrates, $a$ was determined by the cost function gradient $\alpha$ (Fig. \ref{fig_cmn_sim}A), and the power law coefficients of the strength of social relationships increased with the gradient $\alpha$ (Fig. \ref{fig_cmn_sim}B and \ref{fig_cmn_sim}C), that is, the gradient increased the width and shallowness of social relationships. 
Because the increase of the gradient $\alpha$ decreased the number of social relationships of groomers with strong social relationships, they have to invest time in strong social relationships to maintain these relationships.
Consequently, increase in $\alpha$ decreased strong social relationships, and structures of social relationships became relatively wide and shallow.

\section*{Discussion}

There is a trade-off between the number of social relationships (that is, $N$) and the mean strength of social relationships (that is, $m$)~{\bf(Roberts SGB et al., 2009; Miritello G et al., 2013a)} 
as humans must perform frequent social grooming to maintain close relationships~{\bf(Hill RA and Dunbar RIM, 2003; Roberts SGB and Dunbar RIM, 2011; Saramaki J et al., 2014)}. 
Here, we found a simple law where $N$ was inversely proportional to $m^a$ ($a>1$). 
The $a$ is due to the increase in social grooming costs; the costs increase with the strengths of social relationships. 
This cost increase may be due to the fact that strong social relationships tend to be the site of complex and frequent communications. 

We also found that the gradient of the cost increase was an important factor to determine the structure of social relationships. 
In communication systems with the large gradient of social grooming costs, people tend to construct wide and shallow social relationships. 
In contrast, people tend to maintain close social relationships with limited partners in the communication system with a small gradient of social grooming costs.

The variation of the gradient may explain why human beings use different social grooming methods according to their different social relationships. 
Human beings' lightweight social grooming has evolved to adapt large groups, for example, gaze grooming~{\bf(Kobayashi H and Kohshima S, 1997; Kobayashi H and Hashiya K, 2011)} 
and gossip~{\bf(Dunbar RIM, 2004)}, 
because these grooming methods enable humans to have several social relationships and require less time and efforts.
For example, in text communications over the Internet, such as on Twitter, users tend to construct wide and shallow social relationships that are used for acquiring and diffusing information~{\bf(Arnaboldi V et al., 2013c; Preoiuc-Pietro D, 2015)}. 
Thus, lightweight social grooming is effective in constructing many weak social relationships. 
In contrast, elaborate social grooming methods are more effective than lightweight social grooming methods in reinforcing social relationships~{\bf(Burke M and Kraut RE, 2014)}. 
Additionally, for strong social relationships, human beings prefer elaborate social grooming methods, such as face-to-face or telephone communications, rather than text communications over the Internet~{\bf(Burke M and Kraut RE, 2014)}. 
It may be because the degree of satisfaction from elaborate social grooming methods (face to face and video calls) tend to be higher than that from lightweight social grooming methods (phone and text communication)~{\bf(Vlahovic T et al., 2013)}. 
Therefore, we infer that the gradient of elaborate social grooming costs is smaller than that of lightweight social grooming costs, and the intercept of lightweight social grooming methods is smaller than that of elaborate social grooming methods. 
Thereby, human beings tend to construct new social relationships and maintain weak social relationships by using lightweight social grooming methods, and they tend to use elaborate social grooming methods to maintain strong social relationships. 
It thus suggests that an emergence of novel social grooming methods can change social structures, for example, the evolution of gaze grooming~{\bf(Kobayashi H and Kohshima S, 1997; Kobayashi H and Hashiya K, 2011)} 
and gossip~{\bf(Dunbar RIM, 2004)}. 
That is, the evolution of human beings' lightweight social grooming to adapt large groups suggests that they have made wide and shallow societies.
It may shed light on the emergence of our huge modern societies.
Additionally, our findings also suggest predictability of the invention effects of novel communications over the Internet on social structures, that is, in the novel lightweight social grooming will drive wide and shallow social structures, and novel elaborate social grooming will drive narrow and deep social structures.

The strengthening of already strong social relationships (the Yule–-Simon process which generates power law distributions) and the positive gradient of social grooming costs, which represent a time cost distribution strategy, seem to have been caused by competing for cooperating with others. 
That is, strong social relationships may exist for receiving cooperation from others~{\bf(Brown SL and Brown RM, 2006; Miritello G et al., 2013b)}. 
However, cooperators cannot cooperate with everyone because there are costs of cooperation~{\bf(Santos FC, 2006; Xu B and Wang J, 2015)}. 
Actually, human beings tend to cooperate with close friends~{\bf(Haan M et al., 2006; Harrison F et al., 2011; Dunbar RIM, 2016)}. 
Consequently, human beings would need to compete based on the strength of the social relationships with the cooperators' friends and this may generate a skewed distribution for the strength of social relationships. 
Exploring evolutionary stability of this strategy for cooperation will provide knowledge about evolutionary dynamics of human significant social intelligence based on social brain hypothesis.

Additionally, our findings provide novel insight on social network sciences, that is, a constraint on the construction of social relationships ($C=Nm^a$) and its effect on social network structures, for example, they create constraints for the dynamics of network generation and temporal networks.

 \section*{Acknowledgments}
 We are grateful to Assistant Professor Genki Ichinose at Shizuoka University and Mr. Yoshihito Hotta at University of Tokyo, for their valuable comments and suggestions throughout this study.

 \section*{Author contributions statement}

 M.T. designed the research. 
 M.T. conducted data analysis.
 I.F. contributed analysis tools.
 I.F. constructed a data analysis platform for our big data.
 M.T. wrote the main manuscript text. 
 All authors reviewed the manuscript.

\section*{Supplementary Information}

Details of the specification of communication systems (755 and Ameba Pigg).  

\section*{Additional Information}

We declare that this manuscript is original, has not been published before and is not currently being considered for publication elsewhere.
The authors declare no competing financial interests.
All data needed to evaluate the conclusions in the paper are present in the paper, the supplementary information, and DOI: 10.6084/m9.figshare.3395956.v3.

\begin{table}[h]
  \caption{
  Summaries of data sets.
  $N$ and $m$ were tallied for each individual, $d$ was tallied for each relationship and $v$ was tallied for each combination between relationship and day.
}
  \label{tbl_dset}
  \begin{center}
  \footnotesize
    \begin{tabular}{c|c|r|r|r|r|r|r|r|r}
        \hline 
    Communication System & Variable & Size & min & 2.5\%ile & 25\%ile & 50\%ile & 75\%ile & 97.5\%ile & max \\ \hline 
        Twitter & $N$ & 2,585 & 7 & 28 & 65 & 94 & 136 & 264 & 736  \\ 
    & $m$ & 2,585 & 1.25 & 1.84 & 2.79 & 3.55 & 4.66 & 8.61 & 25.23 \\ 
    & $d$ & 278,475 & 1 & 1 & 1 & 1 & 3 & 20 & 166 \\ 
    & $v$ & 943,719 & 2 &   21 & 54 & 94 & 136 & 383 & 14,120 \\ \hline 
    755 Group chat & $N$ & 17,796 &  1  &  1 &  5 & 9 & 17 &  51 & 187\\ 
    & $m$ & 17,796  & 1.00 & 1.43 & 2.44 & 3.53 & 5.60 &  18.00 & 112.00\\ 
    & $d$ & 238,611 &  1 &   1 &  1 & 2 &  4 &  18 & 112\\ 
    & $v$ & 901,212 &  1 &   1 & 17 & 48 & 143 & 1,072 & 31,990\\ \hline 
    755 Wall communication & $N$ & 20,000  &  1 &    1 &   6 & 11 &  24 &  159 & 1,372  \\ 
    & $m$ & 20,000  &  1.00 & 1.02 &  1.53 & 2.45 &  4.39  &  15.67 & 103.00  \\ 
    & $d$ & 534,475 & 1 &   1  & 1 & 1 &  2  & 13 & 121 \\ 
    & $v$ & 1,270,546  & 3 &   6 & 17 & 33 & 73 & 452 & 17,565\\ \hline 
    Ameba Pigg & $N$ & 156,222 & 1  &  1 & 3 & 5 & 13 & 64 & 689 \\ 
    & $m$ & 156,222 & 1.00 & 1.00  & 1.33  & 2.00  & 3.92 & 19.32  & 454.00  \\ 
    & $d$ & 1,911,139 & 1 & 1 & 1 & 1 & 2 & 22 & 457 \\ 
    & $v$ & 6,989,307 & 13 & 143 & 358 & 652 & 1,289 & 4,588 & 87,281\\ \hline 
    Mobile phone & $N$ & 73 &  2  & 16 & 47 & 94 & 126 & 279 & 688 \\ 
    & $m$ & 73 & 1.81 & 2.06 & 3.34 & 3.95 & 4.75 & 7.45 & 8.07 \\
    & $d$ & 7,801 &  1 &   1  & 1 & 1  & 2 &  32 & 207 \\
    & $v$ & 32,728 &  0 &   0 & 24 & 60 & 223 & 10,261 & 328,031\\ \hline 
    SMS& $N$ & 48  & 1 &   1  & 4 &　11 &　19 &　194 &　283  \\ 
    & $m$ & 48  & 1.00 &   1.00 & 1.68 &　2.85 &　4.03 &　11.35 &　30.5  \\ 
    & $d$ & 1,233 &  1 &   1 &  1 & 1  & 2 &　30 &　153   \\
    & $v$ & 4,942 &  1  &  1 &  1 & 3 &  7  & 36 &　168\\ \hline 
    \end{tabular}
  \end{center}
\end{table}

\begin{table}[h]
  \caption{
The results of the regression analysis (Eq. 1) of each communication system.
The t-values and the p-values of $a$ measuring the statistical uncertainty in coefficient $a$ are larger than $1$.
The t-values and the p-values of $b$ measuring the statistical uncertainty in coefficient $b$ are not equal to $0$.
The coefficient $a$ was larger than $1$, that is, the user behaviour data did not obey the null model ($C = Nm$; $a = 1$). 
Their adjusted R-squared values were $0.990$ (Twitter), $0.974$ (755 group chat), $0.959$ (755 wall communication), $0.997$ (Ameba Pigg), $0.994$ (mobile phone) and $0.990$ (SMS).
  }
  \label{tbl_pt_nb}
  \begin{center}
  \footnotesize
    \begin{tabular}{c|c|r|r|r|r}
      \hline         
      Communication System & Coefficient & Estimate & Standard Error & t-value & p-value \\ \hline 
        Twitter & $a$ & $1.189567$ & $0.023256$ & $8.15$ & $4.4 \times 10^{-16}$ \\
                & $b$ & $1.309346$ & $0.006815$ & $192.12$ & Less than $2.0 \times 10^{-16}$ \\ \hline 
        755 Group chat & $a$ & $1.214229$ & $0.004640$ & $46.17$ & Less than $2.0 \times 10^{-16}$ \\
          & $b$ & $1.269766$ & $0.002294$ & $553.5$ & Less than $2.0 \times 10^{-16}$ \\ \hline 
        755 Wall communication & $a$ & $1.562142$ & $0.006250$ & $89.94$ & Less than $2.0 \times 10^{-16}$ \\
         & $b$ & $1.476393$ & $0.002769$ & $533.2$ & Less than $2.0 \times 10^{-16}$ \\ \hline 
        Ameba Pigg & $a$ & $1.0954104$ & $0.0007440$ & $128.24$ & Less than $2.0 \times 10^{-16}$  \\
          & $b$ & $1.0939529$ & $0.0003137$ & $3487$ & Less than $2.0 \times 10^{-16}$ \\ \hline 
        Mobile phone & $a$ & $1.07332$ & $0.15756$ & $0.47$ & $3.2 \times 10^{-1}$ \\ 
                & $b$ & $1.25628$ & $0.04689$ & $26.795$ & Less than $2.0 \times 10^{-16}$ \\ \hline 
        SMS & $a$ & $1.24089$ & $0.07815$ & $3.08 $ & $3.5 \times 10^{-3}$ \\
                & $b$ & $1.21949$ & $0.02995$ & $40.72$ & Less than $2.0\times 10^{-16}$ \\ \hline 

        \end{tabular}

    \end{center}
\end{table}

\end{document}